\documentclass[%
 reprint,
 amsmath,amssymb,
 aps,
]{revtex4-1}
\usepackage{hyperref}
\usepackage{xcolor}
\usepackage{graphicx}
\usepackage{amssymb}
\usepackage{amsmath}
\usepackage{amsfonts}
\usepackage{mathtools}
\usepackage{comment}
\usepackage{tikz}
\usepackage{graphicx}
\usepackage{dcolumn}
\usepackage{bm}

\begin{document}


\def\titlename{Artificial electric field and electron hydrodynamics}
\title{\titlename}
\def\authornames{Omid Tavakol, Yong Baek Kim}
\author{Omid Tavakol}
\author{Yong Baek Kim}


\def\affiliations{Department of Physics, University of Toronto, Toronto, Ontario, M5S 1A7, Canada}
\affiliation{\affiliations}

\begin{abstract}
In the electron dynamics in quantum matter, the Berry curvature of the electronic wave function provides the artificial magnetic field (AMF) in momentum space, which leads to non-trivial contributions to transport coefficients. It is known that in the presence of electron-electron and/or electron-phonon interactions, there is an extra contribution to the electron dynamics due to the artificial electric field (AEF) in momentum space. In this work, we construct hydrodynamic equations for the electrons in time-reversal invariant but inversion-breaking systems and find the novel hydrodynamic coefficients related to the AEF. Furthermore, we investigate the novel linear and non-linear transport coefficients in presence of the AEF.

\end{abstract}


\maketitle

\section{Introduction}
 Transport properties of electrons in quantum matter reflect the nature of the quasi-particle interactions and possible quantum interference effects. The Berry curvature of the electron wave function is the prominent example of the quantum correction to the semi-classical equation of motion of electrons. It stems from a topological property of the electronic wave function in momentum space. Various unusual linear and non-linear transport coefficients have been discussed as the Berry curvature effect in the quasi-particle dynamics, which enters as an artificial magnetic field (AMF) in momentum space \cite{Lecturenoteberry1,Lecturenoteberry2}. Recently, the effect of the AMF on the electron hydrodynamic equations for time-reversal invariant but inversion-symmetry breaking systems is studied in great detail \cite{Kawaki}. For example, it is pointed out that the Poiseuille flow\cite{postulie_current} , is modified in a non-trivial way. Such effects would be of great interest to both high energy and condensed matter physics \cite{hydroLucas}.
 
In the electron hydrodynamics\cite{hydroreviZaanen}, it is assumed that the electron-electron scattering rate $1/\tau_{ee}$ is much greater than other scattering rates such as the electron-phonon $1/\tau_{ep}$ and electron-impurity $1/\tau_{imp}$ scattering rates. The strong electron-electron scattering establishes local equilibrium so that local temperature and chemical potential are well-defined. It is generally hard to achieve this regime in real materials, where typically $1/\tau_{ep}$ ($1/\tau_{imp}$) dominates the high (low) temperature regime. Strongly interacting electrons in ultra-pure systems, however, may offer such a hydrodynamic regime, where a window of temperature exists for $1/\tau_{ee} \gg 1/\tau_{ep}, 1/\tau_{imp}$. Much attention has been paid to graphene, $\text{PdCoO}_2$, and $\text{WP}_2$ as possible candidate materials \cite{MP2,pdco2,Graphine1,Graphine2}. 

In the presence of interactions, it has been known that the semi-classical electron dynamics is affected by the artificial electric field (AEF), which may be regarded as a generalized Berry phase effect in frequency-momentum space \cite{2006Blants}. In addition to the effect of AMF on transport coefficients, we then have to consider the influence of AEF on the electron transport. In the hydrodynamic regime, the momentum relaxation rate is small by definition and it may be considered in the Boltzmann equation via a phenomenological parameter $1/\tau_{rm}$. For example, the small momentum relaxation of electrons may occur due to the weak electron-phonon interactions, which may also be a source of the AEF. \cite{2007Balents} 

In this work, we investigate the electron hydrodynamics by taking into account both AMF and AEF on equal footing. For concreteness, we consider the systems, where time-reversal symmetry is preserved, but the inversion symmetry is broken. We demonstrate that the AEF provides unexpected novel transport and hydrodynamic coefficients. Some explicit examples of the AEF effects on transport and electron hydrodynamics are shown.

 The rest of the paper is organized as follows. In section II, we derive the hydrodynamic equations from the equation of motion and the Boltzmann equation by taking into account both AMF and AEF. In section III, an explicit example of the AEF effect in the presence of a weak electron-phonon interaction is shown and the corresponding transport coefficients are computed. In section IV, we show that the Poiseuille flow becomes fully three-dimensional in the presence of the AEF.


\section{Hydrodynamic equation with AEF}
In this section, we investigate the contribution of AEF in the Boltzmann equation, and its consequences in hydrodynamic coefficients. To do so, we start with the Boltzmann equation in relaxation time approximation. We construct the constitutive relations for stress tensor and momentum to find the hydrodynamic equation for hydrodynamic velocity variable $\vec{u}$. Finally, we find the transport current expressed in terms of hydrodynamic variables and investigate the transport coefficients in a spacially uniform solution.

\subsection{ AEF and equation of motion}
To derive the hydrodynamic equations, we start with semi-classical equations of motions and Boltzmann equation. Both AEF and AMF can be incorporated in the equation of motion
as follows.\cite{2006Blants}
\begin{equation}
\dot{\vec{r}}_{n}=\vec{v}_{n}+(\vec{\Omega}_{n}-\vec{\mathcal{E}}_{n}\times \vec{v}_{n})\times\dot{\vec{k}}_{n} \,,  
\label{1main}
\end{equation}
\begin{equation}
\dot{\vec{k}}_{n}=-e\vec{E} \,.
\label{2main}
\end{equation}
Here, $\vec{E}$ is the external electric field, $n$ is the band index and $\vec{v}_{n}=\frac{\partial \epsilon_{n}(p)}{\partial \vec{p}}$, where $\epsilon_{n}(p)$ is the energy dispersion. $\vec{\Omega}$ and $\vec{\mathcal{E}}$ are the AMF and AEF respectively. For Abeline gauge field $\mathcal{A}^{\alpha}_{\mu}=\big<u_{\alpha}|\frac{\partial}{\partial k_\mu}|u_{\alpha}\big>$ where $|u_{\alpha}\big>$ is the Bloch wave function, AMF and AEF are defined by $\Omega^{\alpha}_{j}=i\epsilon_{jml}\partial_{k_m}\mathcal{A}^{\alpha}_{l}$ and $\mathcal{E}^{\alpha}_{j}=i(\partial_{\omega}\mathcal{A}^{\alpha}_{j}-\partial_{k_j}\mathcal{A}^{\alpha}_{0})$ respectively, where $k_{\mu}=(\omega,\vec{k})$\cite{2007Balents}. Here, $\mathcal{E}_{j}^{\alpha}$ in Eq(\ref{1main}) is evaluated at $\omega=\epsilon_k$.

 The Boltzmann equation that describes the evolution of electron distribution function is given by
\begin{equation}
    \frac{\partial f}{\partial t}+ \dot{\vec{r}}\cdot\nabla_{r}f+\dot{\vec{k}}\cdot\nabla_{k}f=\mathcal{C}[f]\,,
    \label{3}
\end{equation}
where $f(t,r,p)$ is the electron distribution function and $C[f]$ is the collision term. By using Eq(\ref{1main}) and Eq(\ref{2main}) in Eq(\ref{3}) one can find the contribution of AMF and AEF in the Boltzmann equation.
\begin{equation}
       \frac{\partial f}{\partial t}+ \big(\vec{v}+e\vec{E}\times(
   \vec{\Omega}-\vec{\mathcal{E}}\times \vec{v})\big)\cdot\nabla_{r}f-e\vec{E}
   \cdot \nabla_{k}f=\mathcal{C}[f]\,.
   \label{4}
\end{equation}

\subsection{Derivation of hydrodynamic equation}
In the following, we consider the systems, where their band structure near the
Fermi level is constructed of several equivalent valleys with
an isotropic parabolic dispersion with mass $m$ \cite{valleyphysics}. To obtain the hydrodynamic equation for the total momentum, we need to multiply the equation by momentum and integrate over
the momentum space. We consider the collision term  $C[f]=C_{mc}[f]+C_{mr}[f]$ where the first term is related to the collisions that conserve momentum, and the second term is related to the collisions that relax the momentum which we parametrize it with  $\frac{f(t,r,p)}{\tau_{mr}}$ in relaxation time approximation. By integrating over the momentum, the conserved momentum term vanishes, and we can find the following as a hydrodynamic equation. (see the appendix for more details).
\begin{equation}
\frac{\partial \vec{P}}{\partial t} + \vec{\nabla}\cdot\Pi+en\vec{E}=-\frac{\vec{P}}{\tau_{mr}}\,,
\label{5}
\end{equation}
where we can define momentum, stress tensor and density respectively 
\begin{equation}
    \vec{P}=\int [dp]\vec{p}f\,,
    \label{6}
\end{equation}
\begin{equation}
  \Pi_{ij}=\int [dp] p_{i}\big(\vec{v}+eE\times
   \Omega-eE\times (\mathcal{E}\times
   v)\big)_{j}f \,,
   \label{7}
\end{equation}
\begin{equation}
    n=\int [dp]f\,.
    \label{8}
\end{equation}
When there are well defined local temperature and chemical potential, the distribution function $f$ can be written as  $f^{\alpha}_0=\frac{1}{\exp[\beta(\epsilon^{\alpha}(p)-\mu)]+1}$ which is the Fermi-Dirac distribution function, and $[dp]=\int \frac{d^dp}{(2\pi\hbar)^d}$ where $d$ is the spatial dimension.

When we are in the hydrodynamic regime, we can express these quantities in terms of hydrodynamic variables velocity $\vec{u}$, chemical potential $\mu$ and temperature $T$. As a result, in a noncentrosymmetric metals, we find that the presence of the AEF and AMF leads to the
following expressions of the momentum

\begin{equation}
    \vec{P}=\rho \vec{u}\,,
    \label{9}
\end{equation}
and stress tensor 
\begin{equation}
      \Pi_{ij}=\rho u_{i}u_{j}+P \delta_{ij}+e\epsilon_{klj}C_{il}E_{k}+e\mathcal{G}_{kji}E_{k}\,,
      \label{10}
\end{equation}
where $\rho$ is the mass density,  $P$ is the pressure and the coefficients $C_{il}$ and $\mathcal{G}_{ijk}$ are two anomalous coefficients. The $C_{il}$ is reported in \cite{Kawaki} and $\mathcal{G}_{ijk}$ is a novel transport coefficient which is related to the AEF as follows
\begin{equation}
     \mathcal{G}_{ijk}=\sum_{\alpha}\int [dp] (\mathcal{E}^{\alpha}_{i}v^{\alpha}_{j}-\mathcal{E}^{\alpha}_{j}v^{\alpha}_{i})p_{k}f^{\alpha}_{0}\,.
     \label{11}
\end{equation}
The $\mathcal{G}_{ijk}$ is anti-symmetric under exchanging first two indices, $\mathcal{G}_{ijk}=-\mathcal{G}_{ijk}$. Also it is even under time-reversal ( $\mathcal{G}_{ijk}=\mathcal{G}_{ijk}$) and odd under Inversion symmetry ($\mathcal{G}_{ijk}=-\mathcal{G}_{ijk}$). It means, in a system that is invariant under both of these symmetries, $\mathcal{G}_{ijk}$ vanishes.

Knowing the constitutive relation of hydrodynamic quantities, momentum density and stress tensor, we can find the hydrodynamic equation for $\vec{u}$ using Eq(\ref{5})

\begin{align}
&\rho \frac{\partial u_{i}}{\partial t}+\rho u_{j}\partial_{j}u_{i}+\partial_{i} P+em\epsilon_{jkl}E_{k}\big[F_{il}\frac{\partial_{j}T}{T}+D_{il}\partial_{j}\mu\big]\nonumber\\&+e\epsilon_{jkl}C_{il}\partial_{j}E_k
+emE_k\big[I_{jki}\frac{\partial_{j}T}{T}+G_{jki}\partial_{j}\mu\big]\nonumber\\&+e\mathcal{G}_{kji}\partial_jE_k+enE_{i}=-\rho\frac{ u_i}{\tau_{mr}}\,.
\label{12}
\end{align}
Here the transport coefficients $F_{il}$ are $D_{il}$ are from AMF as reported in \cite{Kawaki} and $I_{ijk}$, $G_{ijk}$ are the novel
transport
coefficients resulting from AEF that have the following forms

\begin{equation}
    G_{ijk}=\sum_{\alpha}\int[dp](\mathcal{E}^{\alpha}_{i}v^{\alpha}_{j}-\mathcal{E}^{\alpha}_{j}v^{\alpha}_{i})\frac{\partial f^{\alpha}_{0}}{\partial p_k}\,,
    \label{13}
\end{equation}

\begin{equation}
    I_{ijk}=\sum_{\alpha}\frac{\beta}{m}\int[dp](\mathcal{E}^{\alpha}_{i}v^{\alpha}_{j}-\mathcal{E}^{\alpha}_{j}v^{\alpha}_{i})p_{k}\frac{\partial f^{\alpha}_0}{\partial \beta}\,.
    \label{14}
\end{equation}
These coefficients are related to $\mathcal{G}_{ijk}$ as $G_{ijk}\sim \frac{\partial\mathcal{G}_{ijk}}{\partial \mu}$ and  $I_{ijk}\sim \frac{\partial\mathcal{G}_{ijk}}{\partial T}$ so they have same symmetries as $\mathcal{G}_{ijk}$; both $G_{ijk}$ and $I_{ijk}$ are anti-symmetric tensors under exchanging first two indexes and they are even under time-reversal and odd under Inversion symmetry.

\subsection{Transport current}
One way to investigate system's response to external sources such as electric field  $E=Re[\Tilde{E}e^{i\omega t}]$ and $\nabla T$, is to study the transport current  $\vec{J}$, which is known as\cite{transportcurrent5}\cite{transportcurrent6}
\begin{equation}
    J=\sum_{\alpha}\bigg[e\int[dp]\dot{r}_{\alpha}f_{\alpha}+\nabla \times \int[dp]m_{\alpha}f_{\alpha}\bigg]-\nabla \times M
    \label{15}
\end{equation}
where $M$ is an orbital magnetization. By expanding the terms in hydrodynamic variables we find the following expression for the transport current.

\begin{align}
    J_{i}&=nu_{i}+em\epsilon_{ikl}(E_{k}+\partial_{k}\mu)D_{jl}u_{j}+em\epsilon_{ikl}(\frac{\partial_{k}T}{T})F_{jl}u_{j}\nonumber\\
    &+e\epsilon_{ikl}C_{jl}\partial_{k}u_{j}+emG_{ikj}(E_{k}+\partial_{k}\mu)u_{j}\nonumber\\
    &+emI_{ikj}(\frac{\partial_{k}T}{T})u_{j}-e\mathcal{G}_{ikj}\partial_{k}u_{j}
 \label{16}
\end{align}

As an example, we can look at the uniform solution of Eq(\ref{12}) and find the linear and non-linear transport coefficients in presences of external electric field and temperature gradient. We can define the transport current as $J_i=Re[J^{0}_{i}+J^{\omega}_{i}e^{i\omega t}+J^{2\omega}_{i}e^{2i\omega t}]$ where $J^{\omega}_{i}$ is the linear and $J^{2\omega}_{i}$ is the non-linear current. By considering the uniform solution of Eq(\ref{12}) we can find the on-shell current, we can then find the linear and non-linear transport coefficients. We can write the current as $J=J^{D}+J^{anom}$, where $J^{D}$ is the standard Drude current, and the second term is the anomalous current. We show that in the presence of AEF, there is an additional contribution to $J^{anom}$, which we define as $J^{\mathcal{E}}$. Other contributions to $J^{anom}$ coming from AMF is investigated in \cite{Kawaki}.

\begin{equation}
J^{\mathcal{E}}_{i}=\sigma^{\mathcal{E}}_{ijk}E_{j}E_{k}+\kappa^{\mathcal{E}}_{ijk}\frac{\partial_{j}T}{T}E_{k}+\alpha^{\mathcal{E}}_{ijk}\partial_{j}\mu E_{k} \,,
\end{equation}
where we can define novel transport coefficients as follows.

\begin{equation}
    \sigma^{\mathcal{E}}_{ijk}=\frac{ne^3m}{2(i\omega+\frac{1}{\tau_{mr}})}G_{ijk}\,,
\end{equation}

\begin{equation}
    \kappa^{\mathcal{E}}_{ijk}=\frac{ne^3m}{i\omega+\frac{1}{\tau_{mr}}}\big[I_{ijk}+I_{jki}\big]\,,
\end{equation}

\begin{equation}
    \alpha^{\mathcal{E}}_{ijk}=\frac{ne^3m}{i\omega+\frac{1}{\tau_{mr}}}\big[G_{ijk}+G_{jki}\big]\,.
\end{equation}

All these transport coefficients correspond to non-linear response.


\section{ Effect of AEF on the transport in a two-dimensional
system with electron-phonon interaction
}
In the following, we explain the origin of these transport coefficients
and discuss the consequences by investigating an example. We consider a 2D Hamiltonian model, and we find the AEF due to the electron-phonon interaction. Finally, we investigate the AEF in this model, and find analytic expressions for new transport coefficients in a specific limit.

In the systems with electron-phonon interaction, strictly speaking, we need to consider another Boltzmann equation for phonon distribution function for self-consistency. As mentioned in the \cite{2020phononhydro}, however, phonons in the hydrodynamic regime are much slower than electrons, so most of the contributions to the transport coefficients come from electrons. As a result, we will ignore changes in phonon distribution function and consider them at equilibrium.

\subsection{ Free Hamiltonian}

We consider a 2D system that has two valleys located at
finite momentum positions, $K$ and $K'$. The low-energy effective Hamiltonian near these points is given by \cite{Hamiltonianmodel}:
\begin{equation}
    H^{\alpha}_{0}(\vec{k})=\alpha s k_{y} \mathbf{1}+v k_{x}\tau_{y}-\alpha v k_{y} \tau_{x}+\Delta \tau_{z}\,,
\end{equation}
where $\alpha=\pm$ is related to the valley index. The dispersion relation for this model can be written as 
\begin{equation}
    \epsilon^{\alpha}_{\gamma}(\vec{k})=\alpha s k_{y}+\gamma\lambda(k)\,,
    \label{22}
\end{equation}
where $k=\sqrt{k_{x}^{2}+k_{y}^{2}}$,
 $\lambda(k)=\sqrt{(vk)^2+\Delta^2}$ and $\gamma=\pm$, $\gamma=+$ is for the conduction band and $\gamma=-$ is for the valence band. Also the eigenvectors can be parametrized as 
 
 \begin{equation}
    u^{\alpha}_{+}(\vec{k})=\begin{pmatrix}
cos(\frac{\theta_{k}}{2})\\
sin(\frac{\theta_{k}}{2})e^{i\phi^{\alpha}_{k}}
\end{pmatrix}\,,
\end{equation}

\begin{equation}
    u^{\alpha}_{-}(\vec{k})=\begin{pmatrix}
-sin(\frac{\theta_{k}}{2})e^{-i\phi^{\alpha}_{k}}\\
cos(\frac{\theta_{k}}{2})
\end{pmatrix}\,,
\end{equation}
where

\begin{equation}
    \phi^{\alpha}_{k}=\alpha \phi_{k}+\frac{\pi}{2}\,,
\end{equation}
\begin{equation}
    (cos(\theta_{k}),sin(\theta_{k}))=\frac{1}{\lambda_{k}}(\Delta,v k)\,,
\end{equation}

\begin{equation}
    (cos(\phi_{k}),sin(\phi_{k}))=(\frac{k_{x}}{k},\frac{k_{y}}{k})\,.
\end{equation}

\subsection{ electron-phonon interaction}

\begin{figure}
    \centering
\includegraphics[scale=0.5]{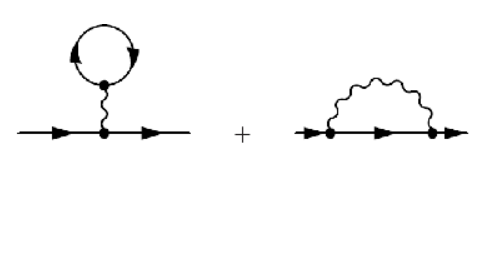}
     \caption{The second order diagrams contributing to the self-energy}
    \label{fig:fymann diagram}
\end{figure}

Now we consider the electron-phonon interaction as the following
\begin{equation}
    H_{ep}=\sum_{k,q,\alpha,\beta}\psi^{\dagger}_{\alpha}(k+q)[g(q)]_{\alpha,\beta}\psi_{\beta}(k)(b_{q}+b^{\dagger}_{-q})
\end{equation}
and 
\begin{equation}
    H_{ph}=\sum_{q}\omega_{0}b^{\dagger}_{q}b_{q}\,,
\end{equation}
where $b_{q}$ is the bosonic field related to phonons, $\omega_{0}$ is a constant frequency, $\psi_{\alpha}(q)$ is the fermionic field related to electrons and $[g(q)]_{\alpha\beta}$ is the electron-phonon coupling. For simplicity we assume $[g(q)]_{\alpha\beta}\approx [g(0)]_{\alpha\beta}$ and because the valleys are located far from each other in k-space, the electron-phonon interaction cannot scatter one electron form a valley to another, then $[g(0)]_{\alpha\beta}\approx g \delta_{\alpha\beta}$. Due to the electron-phonon interaction, the renormalized effective Lagrangian can be written as  $\hat{L}(k,\omega)=H_0(k)+\hat{\sigma}(k,\omega)$ where $\hat{\sigma}(k,\omega)$ is the real part of the self-energy corresponding to the diagrams in Fig.[\ref{fig:fymann diagram}]. which we can write as the following 

\begin{align}
    & \Sigma(k,i\omega_{n})_{\alpha,\beta}=\nonumber\\
     &g^{2}\sum_{m \in even}\int [dq]e^{i\omega_{m}\eta} \mathcal{G}^{(0)}_{\alpha,\beta}(k+q,i\omega_{m})D^{(0)}(q,i\omega_{m}-i\omega_{n})\,,
\end{align}
where $\mathcal{G}^{(0)}(k,i\omega_{n})$  is the free electron propagator 
\begin{equation}
   \mathcal{G}^{(0)}(k,i\omega_{n}) =\sum_{j}\frac{|u_{j}(k)\big>\big<u_{j}(k)|}{i\omega_{n}-(\epsilon_{j}(k)-\mu)}
\end{equation}
and  $D^{(0)}(k,i\omega_{n})$ is the free phonon propagator 
\begin{equation}
    D^{(0)}(k,i\omega_{n})=\frac{1}{i\omega_{n}-\omega_{0}}-\frac{1}{i\omega_{n}+\omega_{0}}\,.
\end{equation}
Here $\omega_{n}$ is the Matsubara frequency, $\eta$ is the small positive number and $|u_{j}(k)\big>$ are the eigenvectors of the $H_{0}$ Hamiltonian. By summing over Matsubara frequencies we can find the following expression for the self energy.
\begin{align}
    \Sigma(k,i\omega_{n})&=\sum_{q}\sum_{\gamma=\pm}|u_{\gamma}(k+q)\big>\big<u_{\gamma}(k+q)|\nonumber\\
&\times \bigg(\frac{n_{f}[\epsilon_{j}(k+q)]+n_{b}[\omega_{0}]}{i\omega_{n}+\omega_{0}-(\epsilon_{j}(k+q)-\mu)}\nonumber\\
&+\frac{n_{b}[\omega_{0}]+1-n_{f}[\epsilon_{j}(k+q)]}{i\omega_{n}-\omega_{0}-(\epsilon_{j}(k+q)-\mu)}\bigg)\,.
\end{align}

\begin{figure}
\includegraphics[scale=0.75]{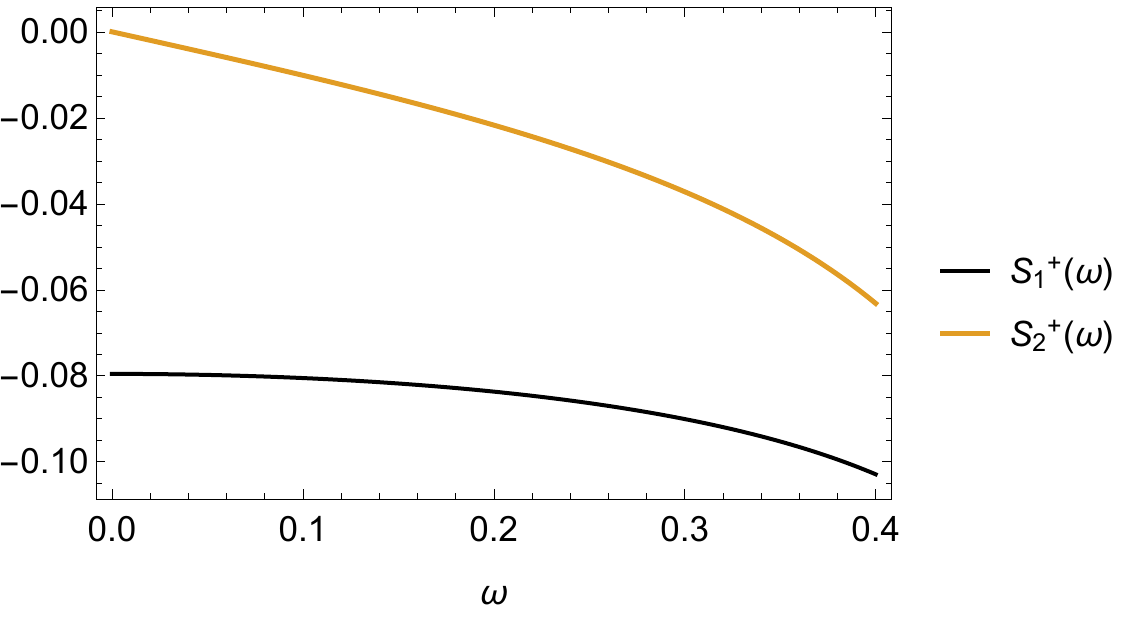}
     \caption{The functions $S_{1}(\omega)$ and $S_{2}(\omega)$ for $\Delta=0.5$, $v=1$, $s=1$, $\omega_0=0.5$ and $g=1$}
    \label{fig2}
\end{figure}

If we analytically continue the imaginary-time self energy, we can find the life-time corresponding to the imaginary part of self-energy and the real part $\hat{\sigma}(k,\omega)$.
In the limit $|\omega-\mu|\ll \omega_0$ and $T \rightarrow 0$, the imaginary part of the self energy vanishes but the real part remains finite even at $T = 0$ \cite{2007Balents}

\begin{align}
    \hat{\sigma}(k,\omega)=&g^{2}\sum_{\gamma=\pm}\bigg[\int_{\epsilon_{\gamma}(k')\leq\mu}\mathcal{P}\frac{|u_{\gamma}(k')\big>\big<u_{\gamma}(k')|}{\omega+\omega_{0}-e_{\gamma}(k')}\nonumber
    \\&+\int_{\epsilon_{\gamma}(k')\geq\mu}\mathcal{P}\frac{|u_{\gamma}(k')\big>\big<u_{\gamma}(k')|}{\omega-\omega_{0}-e_{\gamma}(k')}\bigg]\frac{dk'}{(2\pi)^{2}}\,,
\end{align}
where $e_{\gamma}(k)=\epsilon_{\gamma}(k)-\mu$. By using the results in section A we can write the projection operator as 
\begin{align}
    |u^{\alpha}_{\gamma}(k)\big>&\big<u^{\alpha}_{\gamma}(k)|=\frac{1}{2}\mathbf{1}+\frac{sign(\gamma)}{2}\bigg[cos(\theta_{k})\tau_{z}\nonumber\\
&-\alpha sin(\theta_{k})sin(\phi_{k})\tau_{x}+sin(\theta_{k})cos(\phi_{k})\tau_{y}\bigg]
\end{align}
Using the equation above, we can rewrite $\hat{\sigma}(\omega,k)$ as
\begin{equation}
  \hat{\sigma}^{\alpha}(k,\omega)=S_{0}^{\alpha}(\omega)+S^{\alpha}_{1}(\omega)\tau_{z}-\alpha S^{\alpha}_{2}(\omega)\tau_{x}\,,
\end{equation}
where 

\begin{align}
   S^{\alpha}_{0}(\omega)= &\frac{1}{2}\sum_{\gamma=\pm}\bigg[\int_{\epsilon^{\alpha}_{\gamma}(k)\leq\mu}\mathcal{P}\frac{1}{\omega+\omega_{0}-e^{\alpha}_{\gamma}(k')}\frac{dk'}{(2\pi)^2}\nonumber\\
   &+\int_{\epsilon^{\alpha}_{\gamma}(k)\geq\mu}\mathcal{P}\frac{1}{\omega-\omega_{0}-e^{\alpha}_{\gamma}(k')}\frac{dk'}{(2\pi)^2}\bigg]\,,
\end{align}

\begin{align}
    S^{\alpha}_{1}(\omega)=&\frac{g^2}{2}\sum_{\gamma=\pm}\bigg[\int_{\epsilon^{\alpha}_{\gamma}(k)\leq\mu}\mathcal{P}\frac{sign(\gamma)cos(\theta_{k'})}{\omega+\omega_{0}-e^{\alpha}_{\gamma}(k')}\frac{dk'}{(2\pi)^2}\nonumber\\
    &+\int_{\epsilon^{\alpha}_{\gamma}(k)\geq\mu}\mathcal{P}\frac{sign(\gamma)cos(\theta_{k'})}{\omega-\omega_{0}-e^{\alpha}_{\gamma}(k')}\frac{dk'}{(2\pi)^2}\bigg]\,,
\end{align}
and 
\begin{align}
    &S^{\alpha}_{2}(\omega)=\nonumber\\
    &\frac{g^2}{2}\sum_{\gamma=\pm}\bigg[\int_{\epsilon^{\alpha}_{\gamma}(k)\leq\mu}\mathcal{P}\frac{sign(\gamma)sin(\theta_{k'})sin(\phi_{k'})}{\omega+\omega_{0}-e^{\alpha}_{\gamma}(k')}\frac{dk'}{(2\pi)^2}\nonumber\\
    &+\int_{\epsilon^{\alpha}_{\gamma}(k)\geq\mu}\mathcal{P}\frac{sign(\gamma)sin(\theta_{k'})sin(\phi_{k'})}{\omega-\omega_{0}-e^{\alpha}_{\gamma}(k')}\frac{dk'}{(2\pi)^2}\bigg]\,.
\end{align}
The coefficient of $\tau_{y}$ is zero because the Hamiltonian is invariant under $k_x \rightarrow -k_{x}$ and then the integral is odd under this symmetry. These coefficients can be calculated numerically as shown in Fig.[\ref{fig2}]. 

\subsection{AEF and Transport coefficients}

For any two band systems, we can expand the effective Lagrangian in terms of Pauli matrices 
\begin{equation}
    \hat{L}(k,\omega)=\vec{N}(k,\omega).\vec{\tau}+C(k,\omega)\mathbf{1}\,,
\end{equation}
where 
\begin{equation}
    \vec{N}(k,\omega)=\big(-\alpha\big( v k_{y}+S^{\alpha}_{2}(\omega)\big),\,v k_{x},\,\Delta+S_{1}(\omega)\big)\,,
    \label{42}
\end{equation}
and
\begin{equation}
    C(k,\omega)=\alpha s k_{y}+S_{0}(\omega)\,,
\end{equation}

We use the definition of the AMF and AEF in \cite{2007Balents} to compute these quantities.
\begin{equation}
    \Omega^{\gamma}_{\alpha}(p)=-\frac{sign(\gamma)}{2}(\nabla_{kx}\hat{N}\times\nabla_{ky}\hat{N})\cdot\hat{N}
\end{equation} 
\begin{equation}
 \mathcal{E}_{\gamma,k_{i}}=i\big[(\partial_{\omega}U)^{\dagger}\partial_{k_{i}}U-c.c\big]_{\gamma\gamma}\,,
 \label{45}
\end{equation}
where $U$ is the unitary operator which diagonalizes the effective Lagrangian. 

\begin{equation}
    U= \begin{pmatrix}
cos(\frac{X}{2}) & sin(\frac{X}{2})e^{-iY}\\
-sin(\frac{X}{2})e^{i Y} & cos(\frac{X}{2})
\end{pmatrix}
\label{46}
\end{equation}

and 
\begin{equation}
    X=cos^{-1}\big(\frac{N_{z}}{|N|}\big)\,,
\end{equation}

\begin{equation}
    Y=tan^{-1}\big(\frac{N_{y}}{N_{x}}\big)\,.
\end{equation}
Using Eq(\ref{46}) and Eq(\ref{45}), we can find the following equation for AEF

\begin{equation}
    \mathcal{E}_{\gamma,k_{i}}=sign(\gamma)\bigg((\partial_{\omega}Y)(\partial_{k_{i}}X)-(\partial_{k_{i}}Y)(\partial_{\omega}X)\bigg)\frac{sin(X)}{2}\,.
\end{equation}

Using the definition of $\vec{N}(k,\omega)$ in Eq(\ref{42}), we can find the AEF

\begin{align}
    \big(\mathcal{E}^{\alpha}_{\gamma,x},\mathcal{E}^{\alpha}_{\gamma,y}\big)=&\frac{-sign(\gamma)v^2\alpha}{2(\Delta^2+v^2p^2)^{3/2}}\frac{\partial S^{\alpha}_{1}(\omega)}{\partial \omega}(p_y,-p_x)\nonumber\\
    +&\frac{sign(\gamma)v \Delta \alpha}{2(\Delta^2+v^2p^2)^{3/2}}\frac{\partial S^{\alpha}_{2}(\omega)}{\partial \omega}(1,0)\,.
\end{align}
Also, AMF can be calculated as

\begin{align}
    &\Omega^{\gamma}_{\alpha}(p)=\frac{sign(\gamma)}{2}\frac{\alpha v^2\Delta}{(\Delta^2+v^{2}p^{2})^{3/2}}\nonumber\\
    &+\frac{sign(\gamma)}{2}\frac{\alpha v^2}{(\Delta^2+v^{2}p^{2})^{3/2}}\bigg(\big(1-\frac{3\Delta^2}{(\Delta^2+v^{2}p^{2})}\big)S_{1}(\omega)\bigg)\nonumber\\
    &+\frac{sign(\gamma)}{2}\frac{3k_{y}v^3\alpha\Delta}{(\Delta^2+v^2p^2)^{5/2}}S^{\alpha}_{2}(\omega)\,,
\end{align}
where all the expressions should  be evaluated at $\omega=\epsilon_{\gamma,p}^{(0)}$. AMF is matched with the results in \cite{Kawaki} for a model without electron-phonon interaction. One can see the frequency dependence of the $\vec{N}$ is crucial to have a non-zero AEF and that comes from electron-phonon interaction in our model. To simplify the calculations, we can choose $\mu=0$. Also in weak strain limit, we can approximate the dispersion in Eq(\ref{22}) as $\epsilon^{\alpha}_{\gamma}(k) \sim \frac{(p+p_{\alpha})^2}{2m}+\Delta+O(\frac{s}{v})^2$, where $m=\Delta/v^2$ and $p_{\alpha}=(0,\alpha s \Delta/v^2)$. Finally, we can now estimate the magnitude of the AEF on the Fermi-surface $\omega=\mu$.  

\begin{equation}
   \big(\mathcal{E}^{\alpha}_{\gamma,x},\mathcal{E}^{\alpha}_{\gamma,y}\big)|_{\omega=0}=\frac{sign(\gamma)v^2\Delta}{(\Delta^2+v^2p^2)^{3/2}}g^2sF(1,0) + O(\frac{s}{v})^2\,,
\end{equation}
where 

\begin{equation}
    F=-\int \frac{k^{2}_{y}}{\lambda_{k}(\omega_{0}+\lambda_k)^3}\frac{d^2k}{(2\pi)^2}\,.
\end{equation}

In this model, all the contributions in AEF comes from $S^{\alpha}_{2}(\omega)$ because as one can see in Fig.[\ref{fig2}],  $\frac{\partial S^{\alpha}_{1}(\omega)}{\partial \omega}$ vanishes at $\omega=0$.   
By using these approximations, we can find the analytic expressions for the transport coefficients in section II at $T=0$.  

\begin{equation}
    \mathcal{G}_{xyy}=\frac{g^2s F}{4\pi }\big[-2\Delta+\frac{2\Delta^2+v^2P_{F}^2}{\sqrt{v^2P_{F}^2+\Delta^2}}\big]\,,
\end{equation}
\begin{equation}
    G_{xyy}=2 g^2 s F\frac{P_{F}^2}{(\Delta^2+v^2P_{F}^2)^{3/2}}\,,
\end{equation}
where $  P_{F}=\sqrt{2m|\Delta|}$.


\section{Effect of AEF on
three-dimensional Poiseuille flow
}
In this part, we consider a 3D model with an external electric field in the $y$ direction. The system is bounded in the $x$ direction by the width $w$. To consider boundary effects, the viscosity term is introduced \cite{Kovtun_2012}:

\begin{equation}
    \Pi_{ij}=\rho u_{i}u_{j}+P\delta_{ij}-\Sigma_{ij}\,,
\end{equation}
where
\begin{equation}
    \Sigma_{ij}=\rho \nu(\partial_{i}u_{j}+\partial_{j}u_{i}-\frac{2}{3}\delta_{ij}\partial_{k}u_{k})+\xi\delta_{ij}\partial_{k}u_{k}\,.
\end{equation}
Here $\nu$ is the shear viscosity and $\xi$ is the bulk viscosity, which we are going to ignore.

Now we use an ansatz as a solution, which is  $u_{x}=u_{z}=0$ and $u_{y}=u_{y}(x)$. So the hydrodynamic equation Eq(\ref{12}) becomes:

\begin{equation}
    -\rho \nu \partial^{2}_{x}u_{y}+neE=-\rho\frac{u_{y}}{\tau_{mr}}\,.
\end{equation}
The solution for the above equation with the boundary condition $u(\frac{w}{2})=u(-\frac{w}{2})=0$ is given by:
\begin{equation}
    u_{y}=\frac{e\tau_{mr}E}{m}(1-\frac{cosh(x/l)}{cosh(w/2l)})\,,
\end{equation}
which is a standing wave solution in the $x$ direction,
where 
\begin{equation}
    l=\sqrt{\nu \tau_{mr}}\,,
\end{equation}
and the vorticity is:

\begin{equation}
    \omega_{z}(x)=\frac{\partial u_y}{\partial x}=\frac{e\tau_{mr}E}{ml}\frac{sinh(x/l)}{cosh(w/2l)}\,.
\end{equation}
Now when we have the vorticity, we can compute the on-shell current by using Eq(\ref{16}):
\begin{equation}
    J_{i}=nu_{i}\delta_{i,y}+e\epsilon_{ixl}C_{yl}\partial_{x}u_{y}+emG_{iyy}E_{y}u_{y}-e\mathcal{G}_{ixy}\partial_{x}u_{y}\,.
\end{equation}
As a result we can see that there are contributions in all directions, which come from AEF.
\begin{equation}
J_{x}=-e^2mG_{xyy}Eu_{y}\,,   
\end{equation}
\begin{equation}
J_{y}=-enu_{y}+e^2C_{yz}\omega_{z}+e^2\mathcal{G}_{zxy}\omega_{z} \,,
\end{equation}
\begin{equation}
J_{z}=-e^2C_{yy}\omega_{z}-e^2mG_{zyy}Eu_{y}+e^2\mathcal{G}_{zxy}\omega_{z}\,. 
\end{equation}
The $G_{xyy}$ and $G_{zyy}$ terms are non-linear contributions to the currents, $\mathcal{G}_{zxy}$ and $C_{il}$ terms are linear contributions. \\

\section{ Conclusion and outlook}
We demonstrate that the AEF introduces a number of novel hydrodynamic and non-linear transport coefficients in the time-reversal invariant systems with broken inversion symmetry. In the example of a two-dimensional electron system, we show how these novel transport coefficients arise from the electron-phonon interaction. For instance, it is shown that there is a non-linear transport current along the $x$-direction when the electric field is applied along the $y$-direction, that is $J_x = G_{xyy} E_y E_y$ with a finite $G_{xyy}$. In similar spirit, the Poiseuille flow in the three-dimensional system in a constriction would allow non-uniform (non-linear) transport currents in all three directions. This is in contrast to the usual case, where the non-uniform current exists only along the direction of the applied electric field or even to the case when the AMF effect is included, where there exist the Poiseuille flow in two directions via the presence of the finite vorticity field \cite{Kawaki}.

Our work sheds light on novel non-linear transport and hydrodynamic phenomena in ultra-pure strongly-interacting electron systems. Such systems are great platforms for the discovery of the intricate quantum effects associated with the rather elusive AEF. It will also be interesting to explore further consequences of the AEF in other non-linear electromagnetic responses, both theoretically and experimentally.

\begin{acknowledgments}
This work was supported by the NSERC of Canada. YBK was further supported by the Killam Research Fellowship from the Canada Council for the Arts.
\end{acknowledgments}

\bibliography{bibliography}

\newpage


\appendix
\widetext

\section{Hydrodynamics}

Here we drive the hydrodynamic equation and constitutive relations for the conserved quantities. To find the hydrodynamic relations, we start with modified EoMs:

\begin{equation}
\dot{\vec{r}}_{n}=\vec{v}_{n}+(\vec{\Omega}_{n}-\vec{\mathcal{E}}_{n}\times \vec{v}_{n})\times\dot{\vec{k}}_{n}\,,
\label{1}
\end{equation}

\begin{equation}
\dot{\vec{k}}_{n}=-e\vec{E}\,,
\label{2}
\end{equation}
where $\Omega$ is the Berry curvature, $\mathcal{E}$ is an artificial electric field, $E$ is electric field.
we can see this as a two equation with two unknown variables $\dot{k}_{n}$ and $\dot{r}_{n}$ which need to be solved.

Using $\dot{k}$ in Eq(\ref{2}) and Eq(\ref{1}) we can find:
\begin{equation}
    \dot{\vec{r}}=\vec{v}+eE\times(\vec{\Omega}-\vec{\mathcal{E}}\times \vec{v})\,,   
\end{equation}
When we find $\dot{k}$ and $\dot{r}$ we can write the Boltzmann equation:

\begin{align}
         \frac{\partial f}{\partial t}+ \big(\vec{v}+e\vec{E}\times(
   \vec{\Omega}-\vec{\mathcal{E}}\times \vec{v})\big)\cdot\nabla_{r}f-e\vec{E}
   \cdot \nabla_{k}f=\mathcal{C}[f]\,.
   \label{A4}
\end{align}
We can find hydrodynamic equations by multiplying the above equation by momentum and integrate over
the momentum space that we can find the following equation   

\begin{align}
         \frac{\partial }{\partial t}\int [dk]\vec{k}f&+ \nabla_{r}\cdot\int[dk]\big(\vec{v}+e\vec{E}\times(
   \vec{\Omega}-\vec{\mathcal{E}}\times \vec{v})\big)\vec{k}f-e\vec{E}\int[dk]\vec{k}
   \cdot \nabla_{k}f=\int [dk]\vec{k}\,\mathcal{C}[f]\,,
   \label{A5}
\end{align}
which is the extended hydrodynamic equation for quasi-conserved quantity, momentum.
\begin{equation}
    \frac{\partial \vec{P}}{\partial t} + \vec{\nabla}.\Pi+n\vec{E}=-\frac{\vec{P}}{\tau_{mr}}\,,
    \label{A6}
\end{equation}
where we can define momentum density and modified stress tensor as follows
\begin{equation}
    \vec{P}=\int [dp]\vec{p}f\,,
    \label{A7}
\end{equation}
\begin{equation}
  \Pi_{ij}=\int [dp] p_{i}\big(\vec{v}+e\vec{E}\times
   \vec{\Omega}-e\vec{E}\times (\vec{\mathcal{E}}\times
   \vec{v})_{j}\big)_{j}f  \,.
   \label{A8}
\end{equation}
In the right hand side of the  Eq(\ref{A4}), we consider the collision term  $C[f]=C_{mc}[f]+C_{mr}[f]$ where the first term is related to the collisions that conserve momentum, and the second term is related to the collisions that relax the momentum. So after integration the first term vanishes and we can parametrize the second term with  $\frac{f(t,r,p)}{\tau_{mr}}$ in relaxation time approximation.

To find the constitutive relations for momentum and stress tensor we expand the distribution function in terms of hydrodynamic variables. In the following we assume that the underlying effective theory is invariant under Galilean transformation $f_{0}(\vec{p})=f_{u}(\vec{p}+m\vec{u})$.

To find a relation for the momentum we use Eq(\ref{A7})
\begin{align}
 \vec{P}=&\int [dp]\vec{p}f_{u}(p)=\int [dp](\vec{p}+m\vec{u})f_{0}(p)=mn\vec{u}\,,
\end{align}
where $n=\int[dp]f_0(p)$.
By the same approach we find the constitutive relation for stress tensor using Eq(\ref{A8}), which we rewrite it as
\begin{align}
    \Pi_{ij}&=\sum_{n}\int [dp] p_{i}v_{j,n}f_{u}(p,n)+\int [dp] p_{i}\big(eE\times
   \Omega_n(p)-eE\times (\mathcal{E}_{n}(p)\times
   v_{n})\big)_{j}f_u(p,n)\,,
   \label{A10}
\end{align}
where $n$ is the band index and the first term in the rhs of the above equation is the standard terms for stress tensor in hydrodynamic regime $\int [dp] p_{i}v_{j}=\rho u_iu_j+P\delta_{ij}$. The second term is the anomalous part which we are going to investigate in the following. We denote the first anomalous part as $\Pi^{\Omega}_{ij}$ which means the AMF contributions to the stress tensor.
\begin{align}
     \Pi^{\Omega}_{ij}&=\epsilon_{klj}E_{k}\sum_{n}\int[dp] p_{i}\Omega^{l}_{n}f_{u}\nonumber\\
     &=\epsilon_{klj}E_{k}\int[dp] (p_{i}+mu_{i})\Omega^{l}_{n}(p+m\vec{u})f_{0}(p)\nonumber\\
     &=\epsilon_{klj}E_{k}\int[dp] (p_{i}+mu_{i})\big(\Omega^{l}_{n}+\frac{\partial \Omega^{l}_{n}}{\partial p_{r}}mu_{r}\big)f_{0}(p)\,,
   \label{A10}
\end{align}
up to the second order in $u$ and $E$ we have
\begin{align}
   \Pi^{\Omega}_{ij}=&\sum_{n}\epsilon_{klj}E_{k}\bigg(\int[dp] p_{i}\Omega^{l}_{n}f_{0}+\int[dp] p_{i}\frac{\partial \Omega^{l}_{n}}{\partial p_{r}}f_{0} \,mu_{r}+mu_{i}\int[dp]\Omega^{l}_{n}f_{0}\bigg)\,.
\end{align}
The second term is zero due to time-reversal symmetry($\Omega_{l} \rightarrow -\Omega_{l},p \rightarrow -p$) and the third term is zero because the sum of berry charge over each valley is zero. Finally
\begin{align}
  \Pi^{\Omega}_{ij} =e\epsilon_{klj}E_{k}C_{il}\,,
\end{align}
where 
\begin{equation}
    C_{il}=\sum_{n}\int[dp]p_i\Omega^{l}_{n}f_{0}(p)\,.
\end{equation}

In the following, without loss of generality, we can drop the band index and finally we sum over all bands. For the second part of the anomalous term $\Pi^{\mathcal{E}}_{ij}$ we have

\begin{align}
   &-e \int [dp]p_{i}E\times(\mathcal{E}\times v)_{j}f_{u}=-e\epsilon_{klj}E_{k}\int [dp] p_{i}(\mathcal{E}\times v)_{l}f_{u}\nonumber=-e\epsilon_{klj}\epsilon_{mnl}E_{k}\int [dp] p_{i}\mathcal{E}_{m}v_{n}f_{u}\nonumber\\
  & =-e(\delta_{jm}\delta_{kn}-\delta_{jn}\delta_{km})E_{k}\int [dp] (p_{i}+mu_{i})\mathcal{E}_{m}(p+m\vec{u})v_{n}(p+m\vec{u})f_{0}\nonumber\\
  &=-e(\delta_{jm}\delta_{kn}-\delta_{jn}\delta_{km})E_{k}\int [dp] (p_{i}+mu_{i})(\mathcal{E}_{m}+\frac{\partial \mathcal{E}_{m}}{\partial p_{b}}mu_{b})(v_{n}+\frac{\partial v_{n}}{\partial p_{a}}mu_{a})f_{0}\nonumber\\
  &=-eE_{k}\int [dp] (p_{i}+mu_{i})\bigg[(\mathcal{E}_{j}+\frac{\partial \mathcal{E}_{j}}{\partial p_{b}}mu_{b})(v_{k}+\frac{\partial v_{k}}{\partial p_{a}}mu_{a})-(j\leftrightarrow k)\bigg]f_{0} \Rightarrow\nonumber\\&
 \Pi^{\mathcal{E}}_{ij} =-eE_{k}\bigg(\int [dp]p_{i}\mathcal{E}_{j}v_{k}f_{0}+mu_{i}\int[dp] \mathcal{E}_{j}v_{k}f_{0}+mu_{l}\int[dp] p_{i}v_{k}\frac{\partial \mathcal{E}_{j}}{\partial p_{l}}f_{0}+mu_{l}\int[dp]p_{i}\mathcal{E}_{j}\frac{\partial v_{k}}{\partial p_{l}}f_{0} \bigg)\,.
\end{align}
The first term is odd under Inversion symmetry and even under Time-reversal. The second,third and forth terms are odd under Time-reversal and even under Inversion. So if we consider Time-reversal invariant Noncentrosymmetric system then only the first term is non-zero. Finally we can write the new contribution to the stress tensor as the following
\begin{equation}
    \Pi^{\mathcal{E}}_{ij}=eE_{k}\mathcal{G}_{kji}\,,
\end{equation}
where 
\begin{equation}
    \mathcal{G}_{ijk}=\sum_{n}\int [dp] (\mathcal{E}^{n}_{i}v^{n}_{j}-\mathcal{E}^{n}_{j}v^{n}_{i})p_{k}f^{n}_{0}\,.\label{A17}
\end{equation}

By finding all the contributions we can write the stress tensor as
\begin{equation}
    \Pi_{ij}=\rho u_{i}u_{j}+P \delta_{ij}+e\epsilon_{klj}C_{il}E_{k}+e\mathcal{G}_{kji}E_{k}\,.
\end{equation} 

Now we find the transport current which is made of particle flux $\vec{J}^{N}$ and orbital magnetization.

\begin{equation}
    \vec{J}=\vec{J}^{N}+\nabla\times\vec{M}\,,
    \label{A19}
\end{equation}
where
\begin{equation}
    \vec{J}^{N}=\sum_{n}\int[dp]\dot{\vec{r}}f_{u}=\int[dp](\vec{v}_{n}+e\vec{E}\times(\vec{\Omega}_{n}-\vec{\mathcal{E}}_{n}\times\vec{v}_{n}))f^{n}_u\,
\end{equation}
and 
\begin{equation}
    \vec{M}=\sum_{n}\frac{e}{\beta}\int [dp](\vec{\Omega}_{n}-\vec{\mathcal{E}}_n\times \vec{v}_n)\log(1+e^{-\beta(\epsilon_{n}-\vec{u}\cdot\vec{p}-\mu)})\,.
\end{equation}
Using the Galilean symmetry and expanding up to second order in $u$ and $E$ we find (we should note we drop the band index)
\begin{align}
    J^{N}_{i}=&\int [dp]\bigg[v+\frac{\partial v}{\partial p_{j}}mu_{j}+eE\times \Omega+eE\times\frac{\partial \Omega}{\partial p_{j}}mu_{j}\bigg]f_{0}\nonumber\\&
    -eE\times \int[dp]\bigg[\mathcal{E}\times v+\mathcal{E}\times\frac{\partial v}{\partial p_{j}}mu_{j}+ mu_{j}\frac{\partial \mathcal{E}}{\partial p_{j}}\times v\bigg]f_{0}
    \label{A22}
\end{align}

We separate the terms in particle flux like what we did for stress tensor,
\begin{equation}
    \vec{J}^{N}=\rho \vec{u}+\vec{J}_{\Omega}+\vec{J}_{\mathcal{E}}\,,
\end{equation}
where 
\begin{equation}
    J^{N}_{\Omega,i}=eE_{k}mu_j\epsilon_{kli}\int[dp] \frac{\partial \Omega_{l}}{\partial p_{j}}f_{0}\,.
\end{equation}
Using integrating by part:
\begin{equation}
    J_{\Omega,i}=-eE_{k}mu_j\epsilon_{kli}\int[dp] \Omega_{l}\frac{\partial f_0}{\partial p_{j}}=(em)E_{k}u_{j}\epsilon_{kli}D_{jl}\,,
\end{equation} 
where
 \begin{equation}
     D_{il}=-\sum_{n}\int[dp] \Omega^{n}_{l}\frac{\partial f^{n}_0}{\partial p_{i}}\,.
 \end{equation}

Now for $J^{N}_{\mathcal{E}}$ in Eq(\ref{A22}) we have

\begin{equation}
    J^{N}_{\mathcal{E}}=-eE\times \int[dp]\bigg[\mathcal{E}\times v+\mathcal{E}\times\frac{\partial v}{\partial p_{j}}mu_{j}+ mu_{j}\frac{\partial \mathcal{E}}{\partial p_{j}}\times v\bigg]f_{0}
\end{equation}
If we consider TRS system ($\mathcal{E}\rightarrow\mathcal{E}$, $\vec{p}\rightarrow -\vec{p}$ and $\vec{v}\rightarrow -\vec{v}$)then the first term in $J_{\mathcal{E}}$ vanish.
Using integration by part we find
\begin{equation}
    J_{\mathcal{E},i}=eE_{m}\epsilon_{iml}\epsilon_{nkl}\int[dp]\mathcal{E}_{n}v_{k}\frac{\partial f_{0}}{\partial p_j}mu_j\nonumber
\end{equation}

\begin{align}
   \Rightarrow J_{\mathcal{E},i}=&eE_{k}mu_{j}\int[dp](\mathcal{E}_{i}v_{k}-\mathcal{E}_{k}v_{i})\frac{\partial f_{0}}{\partial p_j}=(em)G_{ikj}E_{k}u_{j}\,,
\end{align}
where
\begin{equation}
    G_{ijk}=\sum_{n}\int[dp](\mathcal{E}^{n}_{i}v^{n}_{j}-\mathcal{E}^{n}_{j}v^{n}_{i})\frac{\partial f^{n}_{0}}{\partial p_k}\,.
\end{equation}

Now for the orbital magnetization part we have the same separation and expansion, $\vec{M}=\vec{M}^{\Omega}+\vec{M}^{\mathcal{E}}$ 
\begin{align}
   & \vec{M}^{\Omega}=\frac{e}{\beta}\int [dp]\vec{\Omega}_{p} log\big(1+e^{-\beta(\epsilon-u.p-\mu)}\big)\nonumber\\&
 \Rightarrow  M^{\Omega}_{i}= eu_l\int[dp]\Omega_ip_lf_{0}+O(u^2)=eu_{l}C_{li}+O(u^2)
\end{align}
\begin{align}
     (\nabla\times \vec{M}^{\Omega})_{k}&=\epsilon_{jik}\partial_{j}M_{i}=e\epsilon_{jik}\partial_{j}(u_{l}C_{li})
     =e\epsilon_{jik}C_{li}\partial_{j}u_{l}+e\epsilon_{jik}u_{l}\partial_{j}C_{li}\,.
\end{align}
We can expand the $\partial_{j}C_{li}$ term:
\begin{align}
\partial_{j}C_{li}=&-\beta\frac{\partial C_{li}}{\partial \beta}(\frac{\partial_{j}T}{T})+\frac{\partial C_{li}}{\partial \mu}\partial_{j}\mu\nonumber\\&=mF_{li}\frac{\partial_{j}T}{T}+mD_{li}\partial_{j}\mu\,,
\end{align}
where
\begin{align}
\frac{\partial C_{li}}{\partial \mu}=&-\int[dp]p_{l}\Omega_{i}\frac{\partial f_{0}}{\partial \epsilon}=m\int[dp]\Omega_{i}\frac{\partial\epsilon}{\partial p_{l}}\frac{\partial f}{\partial \epsilon}\nonumber\\&=m\int[dp]\Omega_{i}\frac{\partial f}{\partial p_{l}}=mD_{li}
\end{align}
and 
\begin{equation}
    F_{il}=\sum_{n}\frac{-\beta}{m}\int[dp]\Omega^{n}_{l}p_{i}\frac{\partial f^{n}_{0}}{\partial \beta}\,.
\end{equation}

 For $\vec{M}^{\mathcal{E}}$ we use similar approach 
\begin{align}
     &\vec{M}_{\mathcal{E}}=-\frac{e}{\beta}\int [dp](\vec{\mathcal{E}}\times \vec{v}) log\big(1+e^{-\beta(\epsilon-u.p-\mu)}\big)\nonumber\\&
  \Rightarrow   \vec{M}^{\mathcal{E}}= -e\int [dp](\vec{\mathcal{E}}\times \vec{v})(\vec{p}\cdot\vec{u}) f_{0}+O(u^2)\,.
\end{align}
Now we can calculate $(\nabla \times \vec{M}_{\mathcal{E}})_{k}$

\begin{align}
    &(\nabla \times \vec{M}_{\mathcal{E}})_{k}=-e\epsilon_{ijk}\partial_{i}\int [dp]\epsilon_{nlj}\mathcal{E}_{n}v_{l}(p.u) f_{0}\nonumber\\
    &= (\delta_{in}\delta_{kl}-\delta_{il}\delta_{kn})\int [dp] \mathcal{E}_{n}v_{l}(p.u)\partial_{i}f_{0}\nonumber\\
    &=\int [dp] (\mathcal{E}_{i}v_{k}-\mathcal{E}_{k}v_{i})(p.u)\big(-\beta\frac{\partial f_{0}}{\partial \beta}(\frac{\partial_{i}T}{T}) +\frac{\partial f_{0}}{\partial \mu}\partial_{i} \mu\big)\nonumber\\&
    +\int [dp] (\mathcal{E}_{i}v_{k}-\mathcal{E}_{k}v_{i})(p_{j}\partial_{i}u_{j})f_{0}\,.
\end{align}
By considering the parabolic dispersion relation
the above equation can be written as the following form
\begin{equation}
    (\nabla\times\vec{M}^{\mathcal{E}})_{k}=I_{kij}(\frac{\partial_{i}T}{T})u_{j}+G_{kij}\partial_{i}\mu u_{j} + \mathcal{G}_{ikj}\partial_{i}u_{j}\,,
\end{equation}
where 
\begin{equation}
    I_{ijk}=\sum_{n}\frac{\beta}{m}\int[dp](\mathcal{E}^{n}_{i}v^{n}_{j}-\mathcal{E}^{n}_{j}v^{n}_{i})p_{k}\frac{\partial f^{n}_0}{\partial \beta}
\end{equation}

Finally we can write the final expression for the transport current Eq(\ref{A19}) 
\begin{align}
    J_{i}&=nu_{i}+em\epsilon_{ikl}(E_{k}+\partial_{k}\mu)D_{jl}u_{j}+em\epsilon_{ikl}(\frac{\partial_{k}T}{T})F_{jl}u_{j}\nonumber\\
    &+\epsilon_{ikl}C_{jl}\partial_{k}u_{j}+emG_{ikj}(E_{k}+\partial_{k}\mu)u_{j}\nonumber\\&+emI_{ikj}(\frac{\partial_{k}T}{T})u_{j}
    -e\mathcal{G}_{ikj}\partial_{k}u_{j}
\end{align}

Using constitutive relations and Eq(\ref{A6}) we can find a hydrodynamic equation

\begin{align}
&\rho \frac{\partial u_{i}}{\partial t}+\rho u_{j}\partial_{j}u_{i}+\partial_{i} P+em\epsilon_{jkl}E_{k}\big[F_{il}\frac{\partial_{j}T}{T}+D_{il}\partial_{j}\mu\big]\nonumber\\&
+e\epsilon_{jkl}C_{il}\partial_{j}E_k
+emE_k\big[I_{jki}\frac{\partial_{j}T}{T}+G_{jki}\partial_{j}\mu\big]\nonumber\\&
+e\mathcal{G}_{kji}\partial_jE_k+enE_{i}=-\rho\frac{ u_i}{\tau_{mr}}\,,
\end{align}

where we used following relation for coefficients 
\begin{align}
    \partial_{j}\mathcal{G}_{kji}=\frac{\mathcal{G}_{kji}}{\partial T}\partial_{j}T+\frac{\mathcal{G}_{kji}}{\partial \mu}\partial_{j}\mu=mI_{jki}\frac{\partial_{j}T}{T}+mG_{jki}\partial_{j}\mu\,.
\end{align}


\section{The model}
Here, we outline the calculation of transport coefficients in a specific Hamiltonian model
\begin{equation}
    H=H^{\alpha}_0+H_{ph}+H_{ep}\,,
\end{equation}
where
\begin{equation}
    H^{\alpha}_{0}(\vec{k})=\alpha s k_{y} \mathbf{1}+v k_{x}\tau_{y}-\alpha v k_{y} \tau_{x}+\Delta \tau_{z}\,,
\end{equation}
\begin{equation}
    H_{ph}=\sum_{q}\omega_{0}b^{\dagger}_{q}b_{q}
\end{equation}
and
\begin{equation}
   H_{ep}=g\sum_{k,q,\alpha}\psi^{\dagger}_{\alpha}(k+q)\psi_{\alpha}(k)(b_{q}+b^{\dagger}_{-q})\,.
\end{equation}
Using perturbation theory we have the following expression for the first order correction to the green's function
\begin{equation}
    \mathcal{G}^{(1)}(k,i\omega_{n})=\mathcal{G}^{(0)}\Sigma(k,i\omega_n) \,\mathcal{G}^{(0)}\,,
\end{equation}
where we can define the self energy as 

\begin{equation}
        \Sigma(k,i\omega_{n})_{\alpha,\beta}\nonumber=g^{2}\sum_{m \in odd}\int [dq]e^{i\omega_{m}\eta} \mathcal{G}^{(0)}_{\alpha,\beta}(k+q,i\omega_{m})D^{(0)}(q,i\omega_{m}-i\omega_{n})\label{AIdonknow}
\end{equation}
$g^{(0)}$ and $D^{(0)}$ are defined by the free electron and phonon's Hamiltonian:

\begin{equation}
   \mathcal{G}^{(0)}(k,i\omega_{n})=\sum_{\gamma}\frac{|u^{\alpha}_{\gamma}(k)\big>\big<u^{\alpha}_{\gamma}(k)|}{i\omega_{n}-(\epsilon^{\alpha}_{\gamma}(k)-\mu)}\,,\label{B6}
\end{equation}

\begin{equation}
    D^{(0)}(k,i\omega_{n})=\frac{1}{i\omega_{n}-\omega_{0}}-\frac{1}{i\omega_{n}+\omega_{0}}\label{B7}\,,
\end{equation}
where $\epsilon^{\alpha}_{\gamma}(\vec{k})=\alpha s k_{y}+\gamma\lambda(k)$ is the eigenvalue of the $H_0$ 
, $k=\sqrt{k_{x}^{2}+k_{y}^{2}}$
and  $\lambda(k)=\sqrt{(vk)^2+\Delta^2}$.
Also for the eigenvectors we have 
\begin{equation}
    |u^{\alpha}_{+}(\vec{k})\big>=\begin{pmatrix}
cos(\frac{\theta_{k}}{2})\\
sin(\frac{\theta_{k}}{2})e^{i\phi^{\alpha}_{k}}
\end{pmatrix}\,,
\end{equation}

\begin{equation}
    |u^{\alpha}_{-}(\vec{k})\big>=\begin{pmatrix}
-sin(\frac{\theta_{k}}{2})e^{-i\phi^{\alpha}_{k}}\\
cos(\frac{\theta_{k}}{2})
\end{pmatrix}\,,
\end{equation}
where

\begin{equation}
    \phi^{\alpha}_{k}=\alpha \phi_{k}+\frac{\pi}{2}\,,
\end{equation}
\begin{equation}
    (cos(\theta_{k}),sin(\theta_{k}))=\frac{1}{\lambda_{k}}(\Delta,v k)
\end{equation}
and

\begin{equation}
    (cos(\phi_{k}),sin(\phi_{k}))=(\frac{k_{x}}{k},\frac{k_{y}}{k})\,.
\end{equation}

Using Eq(\ref{AIdonknow}), Eq(\ref{B6}) and Eq(\ref{B7}) we can find

\begin{align}
    \Sigma^{\alpha}(k,i\omega_{n})=\sum_{m \in odd}\int [dq]e^{i\omega_{m}\eta}\sum_{\gamma}\frac{|u^{\alpha}_{\gamma}(k+q)\big>\big<u^{\alpha}_{\gamma}(k+q)|}{i\omega_{m}-(\epsilon^{\alpha}_{\gamma}(k+q)-\mu)}\big(\frac{1}{i\omega_{m}-i\omega_{n}-\omega_{0}}-\frac{1}{i\omega_{m}-i\omega_{n}+\omega_{0}}\big)
    \label{B13}
\end{align}

where $\eta$ is a small positive number. By calculating the following expression

\begin{align}
   &\sum_{m \in odd}e^{i\omega_{m}\eta}\frac{1}{i\omega_{m}-e^{\alpha}_{\gamma}(k+q)}\frac{1}{i\omega_{m}-i\omega_{n}-\omega_{0}}\nonumber\\&
=\frac{1}{i\omega_{n}+\omega_{0}-e^{\alpha}_{\gamma}(k+q)}\sum_{m \in even}e^{i\omega_{m}\eta}\big[\frac{1}{i\omega_{m}-i\omega_{n}-\omega_{0}}-\frac{1}{i\omega_{m}-e^{\alpha}_{\gamma}(k+q)}\big]\nonumber\\&
=\frac{1}{i\omega_{n}+\omega_{0}-e^{\alpha}_{\gamma}(k+q)}\big(n_{f}[e^{\alpha}_{\gamma}(k+q)]+n_{b}[\omega_{0}]\big)\,,
\end{align}

and summing over Matsubara frequencies in Eq(\ref{B13}) we can find 

\begin{align}
 \Sigma(k,i\omega_{n})&=\sum_{q}\sum_{\gamma=\pm}|u_{\gamma}(k+q)\big>\big<u_{\gamma}(k+q)| \bigg(\frac{n_{f}[e_{j}(k+q)]+n_{b}[\omega_{0}]}{i\omega_{n}+\omega_{0}-e_{j}(k+q)}+\frac{n_{b}[\omega_{0}]+1-n_{f}[e_{j}(k+q)]}{i\omega_{n}-\omega_{0}-e_{j}(k+q)}\bigg)\,.
\end{align}
Using analytic continuation,
in the limit $|\omega-\mu|\ll \omega_0$ and $T \rightarrow 0$, we can find the real part of the self energy as the following 

\begin{align}
    \sigma(k,\omega)=&g^{2}\sum_{\gamma=\pm}\bigg[\int_{\epsilon_{\gamma}(k')\leq\mu}\mathcal{P}\frac{|u_{\gamma}(k')\big>\big<u_{\gamma}(k')|}{\omega+\omega_{0}-e_{\gamma}(k')}
    +\int_{\epsilon_{\gamma}(k')\geq\mu}\mathcal{P}\frac{|u_{\gamma}(k')\big>\big<u_{\gamma}(k')|}{\omega-\omega_{0}-e_{\gamma}(k')}\bigg]\frac{dk'}{(2\pi)^{2}}\,.
    \label{B16}
\end{align}
The projection operator for the mentioned model is 
\begin{align}
    |u^{\alpha}_{\gamma}(k)\big>&\big<u^{\alpha}_{\gamma}(k)|=\frac{1}{2}\mathbf{1}\nonumber+\frac{sign(\gamma)}{2}\bigg[cos(\theta_{k})\tau_{z}-\alpha sin(\theta_{k})sin(\phi_{k})\tau_{x}+sin(\theta_{k})cos(\phi_{k})\tau_{y}\bigg]\,.
\end{align}
 Using projection operator in Eq(\ref{B16}) we can find 

\begin{align}
    \sigma^{\alpha}(k,\omega)&=\frac{1}{2}\sum_{\gamma=\pm}\bigg[\int_{\epsilon^{\alpha}_{\gamma}(k)\leq\mu}\mathcal{P}\frac{1}{\omega+\omega_{0}-e^{\alpha}_{\gamma}(k')}\frac{dk'}{(2\pi)^2}+\int_{\epsilon^{\alpha}_{\gamma}(k)\geq\mu}\mathcal{P}\frac{1}{\omega-\omega_{0}-e^{\alpha}_{\gamma}(k')}\frac{dk'}{(2\pi)^2}\bigg]\mathbf{1}
\nonumber\\&+\tau_{z}\frac{g^2}{2}\sum_{\gamma=\pm}\bigg[\int_{\epsilon^{\alpha}_{\gamma}(k)\leq\mu}\mathcal{P}\frac{sign(\gamma)cos(\theta_{k'})}{\omega+\omega_{0}-e^{\alpha}_{\gamma}(k')}\frac{dk'}{(2\pi)^2}+\int_{\epsilon^{\alpha}_{\gamma}(k)\geq\mu}\mathcal{P}\frac{sign(\gamma)cos(\theta_{k'})}{\omega-\omega_{0}-e^{\alpha}_{\gamma}(k')}\frac{dk'}{(2\pi)^2}\bigg]
\nonumber\\&-\alpha \tau_{x}\frac{g^2}{2}\sum_{\gamma=\pm}\bigg[\int_{\epsilon^{\alpha}_{\gamma}(k)\leq\mu}\mathcal{P}\frac{sign(\gamma)sin(\theta_{k'})sin(\phi_{k'})}{\omega+\omega_{0}-e^{\alpha}_{\gamma}(k')}\frac{dk'}{(2\pi)^2}+\int_{\epsilon^{\alpha}_{\gamma}(k)\geq\mu}\mathcal{P}\frac{sign(\gamma)sin(\theta_{k'})sin(\phi_{k'})}{\omega-\omega_{0}-e^{\alpha}_{\gamma}(k')}\frac{dk'}{(2\pi)^2}\bigg]\,.
\end{align}

The coefficient of $\tau_{y}$ term vanishes because it is an odd function on $k'_x$. Now we can write the above equation in a simpler form
\begin{equation}
    \hat{\sigma}^{\alpha}(k,\omega)=S_{0}^{\alpha}(\omega)+S^{\alpha}_{1}(\omega)\tau_{z}-\alpha S^{\alpha}_{2}(\omega)\tau_{x}
\end{equation}
where 
\begin{align}
   S^{\alpha}_{0}(\omega)= \frac{1}{2}\sum_{\gamma=\pm}\bigg[\int_{\epsilon^{\alpha}_{\gamma}(k)\leq\mu}\mathcal{P}\frac{1}{\omega+\omega_{0}-e^{\alpha}_{\gamma}(k')}\frac{dk'}{(2\pi)^2}
   +\int_{\epsilon^{\alpha}_{\gamma}(k)\geq\mu}\mathcal{P}\frac{1}{\omega-\omega_{0}-e^{\alpha}_{\gamma}(k')}\frac{dk'}{(2\pi)^2}\bigg]\,,
\end{align}

\begin{align}
    S^{\alpha}_{1}(\omega)=\frac{g^2}{2}\sum_{\gamma=\pm}\bigg[\int_{\epsilon^{\alpha}_{\gamma}(k)\leq\mu}\mathcal{P}\frac{sign(\gamma)cos(\theta_{k'})}{\omega+\omega_{0}-e^{\alpha}_{\gamma}(k')}\frac{dk'}{(2\pi)^2}
    +\int_{\epsilon^{\alpha}_{\gamma}(k)\geq\mu}\mathcal{P}\frac{sign(\gamma)cos(\theta_{k'})}{\omega-\omega_{0}-e^{\alpha}_{\gamma}(k')}\frac{dk'}{(2\pi)^2}\bigg]\,,
\end{align}
and 
\begin{align}
    S^{\alpha}_{2}(\omega)=
    &\frac{g^2}{2}\sum_{\gamma=\pm}\bigg[\int_{\epsilon^{\alpha}_{\gamma}(k)\leq\mu}\mathcal{P}\frac{sign(\gamma)sin(\theta_{k'})sin(\phi_{k'})}{\omega+\omega_{0}-e^{\alpha}_{\gamma}(k')}\frac{dk'}{(2\pi)^2}
    +\int_{\epsilon^{\alpha}_{\gamma}(k)\geq\mu}\mathcal{P}\frac{sign(\gamma)sin(\theta_{k'})sin(\phi_{k'})}{\omega-\omega_{0}-e^{\alpha}_{\gamma}(k')}\frac{dk'}{(2\pi)^2}\bigg]\,.
\end{align}
Now one can write the effective Lagrangian $\hat{L}(k,\omega)=H_{0}(k)+\hat{\sigma}(k,\omega)$ as an expansion of Pauli matrices $\hat{L}(k,\omega)=N_{\mu}\tau_{\mu}+C(k,\omega)\mathbf{1}$

\begin{equation}
    \hat{L}(k,\omega)=-\alpha\big( v k_{y}+S^{\alpha}_{2}(\omega)\big) \tau_{x}+v k_{x}\tau_{y}+(\Delta+S_{1}(\omega)) \tau_{z}+(\alpha s k_{y}+S_{0}(\omega)) \mathbf{1}\,,
\end{equation}
where
\begin{equation}
    N_{\mu}=\big(-\alpha\big( v k_{y}+S^{\alpha}_{2}(\omega)\big),\,v k_{x},\,\Delta+S_{1}(\omega)\big)\,.
\end{equation}

To find the AMF we use following definition
\begin{align}
    &\Omega^{\gamma}_{\alpha}(p)=-\frac{sign(\gamma)}{2}(\nabla_{kx}\hat{N}\times\nabla_{ky}\hat{N}).\hat{N}\nonumber\\&
    \Rightarrow \Omega^{\gamma}_{\alpha}(p)=\frac{sign(\gamma)}{2}\frac{\alpha v^2}{(\Delta^2+v^{2}p^{2})^{3/2}}
    \bigg(\Delta+\big(1-\frac{3\Delta^2}{(\Delta^2+v^{2}p^{2})}\big)S_{1}(\omega)|_{\omega=\epsilon_{\gamma,p}^{(0)}}\bigg)\frac{sign(\gamma)}{2}\frac{3k_{y}v^3\alpha\Delta}{(\Delta^2+v^2p^2)^{5/2}}S^{\alpha}_{2}(\omega)
\end{align}

The second term is the correction to the AMF up to second order in electron-phonon coupling .

To compute AEF we need to find an unitary operator ($U$) that diagonalize $\hat{L(k,\omega)}$ and then we can define AEF as:

\begin{equation}
    \mathcal{E}_{\gamma,k_{i}}=i\big[(\partial_{\omega}U)^{\dagger}\partial_{k_{i}}U-c.c\big]_{\gamma\gamma}\,,
\end{equation}

Where

\begin{equation}
    U= \begin{pmatrix}
cos(\frac{X}{2}) & sin(\frac{X}{2})e^{-iY}\\
-sin(\frac{X}{2})e^{i Y} & cos(\frac{X}{2})
\end{pmatrix}
\end{equation}

and 
\begin{equation}
    X=cos^{-1}\big(\frac{N_{z}}{|N|}\big)\,,
\end{equation}

\begin{equation}
    Y=tan^{-1}\big(\frac{N_{y}}{N_{x}}\big)\,.
\end{equation}
By this parametrization the AEF is given by

\begin{equation}
    \mathcal{E}_{\gamma,k_{i}}=sign(\gamma)\bigg((\partial_{\omega}Y)(\partial_{k_{i}}X)-(\partial_{k_{i}}Y)(\partial_{\omega}X)\bigg)\frac{sin(X)}{2}
\end{equation}

finally for this model the AEF becomes:
\begin{align}
    \big(\mathcal{E}^{\alpha}_{\gamma,x},\mathcal{E}^{\alpha}_{\gamma,y}\big)=&\frac{-sign(\gamma)v^2\alpha}{2(\Delta^2+v^2p^2)^{3/2}}\frac{\partial S^{\alpha}_{1}(\omega)}{\partial \omega}(p_y,-p_x)+\frac{sign(\gamma)v \Delta \alpha}{2(\Delta^2+v^2p^2)^{3/2}}\frac{\partial S^{\alpha}_{2}(\omega)}{\partial \omega}(1,0)
\end{align}
we define the notation $\epsilon^{\alpha}_{p}\equiv\epsilon^{\alpha}_{+}$ and $\epsilon^{\alpha}_{m}\equiv\epsilon^{\alpha}_{-}$. Also it is useful to mention some relation between energy dispersion in this model:

\begin{equation}
    \epsilon^{-}_{m}=-\epsilon^{+}_{p}=-(s k_{y}+\lambda(k))
    \label{23}
\end{equation}

\begin{equation}
    \epsilon^{-}_{p}=-\epsilon^{+}_{m}=-(s k_{y}-\lambda(k))
    \label{24}
\end{equation}
Now we calculate $\frac{\partial S^{\alpha}_{1}(\omega)}{\partial\omega}$ and $\frac{\partial S^{\alpha}_{2}(\omega)}{\partial\omega}$ in the following 

\[ \frac{\partial S^{\alpha}_{1}(\omega)}{\partial \omega}|_{\omega=0}=-\frac{g^2}{2}\sum_{\gamma=\pm}\bigg[\int_{\epsilon^{\alpha}_{\gamma}(k)\leq0}\mathcal{P}\frac{sign(\gamma)cos(\theta_k')}{(\omega_{0}-\epsilon^{\alpha}_{\gamma}(k'))^2}\frac{dk'}{(2\pi)^2}+\int_{\epsilon^{\alpha}_{\gamma}(k)\geq0}\mathcal{P}\frac{sign(\gamma)cos(\theta_k')}{(\omega_{0}+\epsilon^{\alpha}_{\gamma}(k'))^2}\frac{dk'}{(2\pi)^2}\bigg]\]

for $\alpha=+$ we have:

\begin{align} 
 \frac{\partial S^{+}_{1}(\omega)}{\partial \omega}|_{\omega=0}&=-\frac{g^2}{2}\bigg[\int_{\epsilon^{+}_{p}(k)\leq0}\mathcal{P}\frac{cos(\theta_k')}{(\omega_{0}-\epsilon^{+}_{p}(k'))^2}\frac{dk'}{(2\pi)^2}-\int_{\epsilon^{+}_{m}(k)\leq0}\mathcal{P}\frac{cos(\theta_k')}{(\omega_{0}-\epsilon^{+}_{m}(k'))^2}\frac{dk'}{(2\pi)^2}\bigg]\nonumber\\&
-\frac{g^2}{2}\bigg[\int_{\epsilon^{+}_{p}(k)\geq0}\mathcal{P}\frac{cos(\theta_k')}{(\omega_{0}+\epsilon^{+}_{p}(k'))^2}\frac{dk'}{(2\pi)^2}-\int_{\epsilon^{+}_{m}(k)\geq0}\mathcal{P}\frac{cos(\theta_k')}{(\omega_{0}+\epsilon^{+}_{m}(k'))^2}\frac{dk'}{(2\pi)^2}\bigg]
\end{align}

for $\alpha=-$ we have:

\begin{align}
    \frac{\partial S^{-}_{1}(\omega)}{\partial \omega}|_{\omega=0}&=-\frac{g^2}{2}\bigg[\int_{\epsilon^{-}_{p}(k)\leq0}\mathcal{P}\frac{cos(\theta_k')}{(\omega_{0}-\epsilon^{-}_{p}(k'))^2}\frac{dk'}{(2\pi)^2}-\int_{\epsilon^{-}_{m}(k)\leq0}\mathcal{P}\frac{cos(\theta_k')}{(\omega_{0}-\epsilon^{-}_{m}(k'))^2}\frac{dk'}{(2\pi)^2}\bigg]\nonumber\\&
-\frac{g^2}{2}\bigg[\int_{\epsilon^{-}_{p}(k)\geq0}\mathcal{P}\frac{cos(\theta_k')}{(\omega_{0}+\epsilon^{-}_{p}(k'))^2}\frac{dk'}{(2\pi)^2}-\int_{\epsilon^{-}_{m}(k)\geq0}\mathcal{P}\frac{cos(\theta_k')}{(\omega_{0}+\epsilon^{-}_{m}(k'))^2}\frac{dk'}{(2\pi)^2}\bigg]
\end{align}
Using eq(\ref{23}) and eq(\ref{24}) we find:

\begin{align}  
\frac{\partial S^{-}_{1}(\omega)}{\partial \omega}|_{\omega=0}&=-\frac{g^2}{2}\bigg[\int_{\epsilon^{+}_{m}(k)\geq0}\mathcal{P}\frac{cos(\theta_k')}{(\omega_{0}+\epsilon^{+}_{m}(k'))^2}\frac{dk'}{(2\pi)^2}-\int_{\epsilon^{+}_{p}(k)\geq0}\mathcal{P}\frac{cos(\theta_k')}{(\omega_{0}+\epsilon^{+}_{p}(k'))^2}\frac{dk'}{(2\pi)^2}\bigg]\nonumber\\&
-\frac{g^2}{2}\bigg[\int_{\epsilon^{+}_{m}(k)\leq0}\mathcal{P}\frac{cos(\theta_k')}{(\omega_{0}-\epsilon^{+}_{m}(k'))^2}\frac{dk'}{(2\pi)^2}-\int_{\epsilon^{+}_{p}(k)\leq0}\mathcal{P}\frac{cos(\theta_k')}{(\omega_{0}-\epsilon^{+}_{p}(k'))^2}\frac{dk'}{(2\pi)^2}\bigg]
\end{align}
$ \frac{\partial S^{-}_{1}(\omega)}{\partial \omega}|_{\omega=0}=- \frac{\partial S^{+}_{1}(\omega)}{\partial \omega}|_{\omega=0}$ is an odd function of $\alpha$. Also the numerical result shows $\frac{\partial S^{\alpha}_{1}(\omega)}{\partial \omega}|_{\omega=0}$ vanishes at $\omega=0$
 
 Now we are going to calculate  $\frac{\partial S^{+}_{2}(\omega)}{\partial \omega}|_{\omega=0}$

\[ \frac{\partial S^{\alpha}_{2}(\omega)}{\partial \omega}|_{\omega=0}=\alpha\frac{g^2}{2}\sum_{\gamma=\pm}\bigg[\int_{\epsilon^{\alpha}_{\gamma}(k)\leq0}\mathcal{P}\frac{sign(\gamma)sin(\theta_{k'})sin(\phi_{k'})}{(\omega_{0}-\epsilon^{\alpha}_{\gamma}(k'))^2}\frac{dk'}{(2\pi)^2}+\int_{\epsilon^{\alpha}_{\gamma}(k)\geq0}\mathcal{P}\frac{sign(\gamma)sin(\theta_{k'})sin(\phi_{k'})}{(\omega_{0}+\epsilon^{\alpha}_{\gamma}(k'))^2}\frac{dk'}{(2\pi)^2}\bigg]\]

for $\alpha=+$:

\[ \frac{\partial S^{+}_{2}(\omega)}{\partial \omega}|_{\omega=0}=\frac{g^2}{2}\bigg[\int_{\epsilon^{+}_{p}(k)\leq0}\mathcal{P}\frac{sin(\theta_{k'})sin(\phi_{k'})}{(\omega_{0}-\epsilon^{+}_{p}(k'))^2}\frac{dk'}{(2\pi)^2}
-\int_{\epsilon^{+}_{m}(k)\leq0}\mathcal{P}\frac{sin(\theta_{k'})sin(\phi_{k'})}{(\omega_{0}-\epsilon^{+}_{m}(k'))^2}\frac{dk'}{(2\pi)^2}\bigg]\]
\[+\frac{g^2}{2}\bigg[\int_{\epsilon^{+}_{p}(k)\geq0}\mathcal{P}\frac{sin(\theta_{k'})sin(\phi_{k'})}{(\omega_{0}+\epsilon^{+}_{p}(k'))^2}\frac{dk'}{(2\pi)^2}
-\int_{\epsilon^{+}_{m}(k)\geq0}\mathcal{P}\frac{sin(\theta_{k'})sin(\phi_{k'})}{(\omega_{0}+\epsilon^{+}_{m}(k'))^2}\frac{dk'}{(2\pi)^2}\bigg]\,.\]

If we assume that $\frac{s}{v}\ll1$ then we can approximate $\epsilon(p)\approx \frac{(p+p_{\alpha})^2}{2m}+\Delta+O(\frac{s}{v})^2$ so we can expand the function to the first order in $(\frac{s}{v})$

\begin{equation}
\frac{\partial S^{\alpha}_{2}(\omega)}{\partial \omega}|_{\omega=0}=\sum_{\gamma=\pm}\frac{g^2}{2}\bigg[\int_{\epsilon^{\alpha}_{\gamma}\leq 0}\frac{sing(\gamma)sin(\theta_k)sin(\phi_k)}{(\omega_{0}-\alpha s k_y-sign(\gamma)\lambda_k)^2}\frac{d^2k}{(2\pi)^2}
\end{equation}
\[+\int_{\epsilon^{\alpha}_{\gamma}\geq 0}\frac{sing(\gamma)sin(\theta_k)sin(\phi_k)}{(\omega_{0}+\alpha s k_y+sign(\gamma)\lambda_k)^2}\frac{d^2k}{(2\pi)^2} \bigg]\]

summing over $\gamma$.
\begin{equation}
  \frac{\partial S^{\alpha}_{2}(\omega)}{\partial \omega}|_{\omega=0}=  \frac{g^2}{2}\bigg[\int\frac{-sin(\theta_k)sin(\phi_k)}{(\omega_{0}-\alpha s k_y+\lambda_k)^2}\frac{d^2k}{(2\pi)^2}+\int\frac{sin(\theta_k)sin(\phi_k)}{(\omega_{0}+\alpha s k_y+\lambda_k)^2}\frac{d^2k}{(2\pi)^2} \bigg]\,,
\end{equation}
Also we expand the denominator of the above expression up to the first order in $(\frac{s}{v})$
\begin{equation}
    (\omega_{0}\pm \alpha s k_y+\lambda_k)^{-2}\approx\frac{1}{ (\omega_{0}+\lambda_k)^{2}}(1\mp\frac{2\alpha s k_y}{\omega_{0}+\lambda_k} )
\end{equation}
which we can find 
\begin{equation}
    \Rightarrow \frac{\partial S^{\alpha}_{2}(\omega)}{\partial \omega}|_{\omega=0}=-2g^2\alpha s \int \frac{vk^{2}_{y}}{\lambda_{k}(\omega_{0}+\lambda_k)^3}\frac{d^2k}{(2\pi)^2}
\end{equation}
and by defining $  F=-\int \frac{k^{2}_{y}}{\lambda_{k}(\omega_{0}+\lambda_k)^3}\frac{d^2k}{(2\pi)^2}$ we have 
\begin{equation}
    \frac{\partial S^{\alpha}_{2}(\omega)}{\partial \omega}|_{\omega=0}=2g^2\alpha s v F\,.
    \label{A39}
\end{equation}
Finally we find the AEF as:

\begin{equation}
   \big(\mathcal{E}^{\alpha}_{\gamma,x},\mathcal{E}^{\alpha}_{\gamma,y}\big)=\frac{sign(\gamma)v^2\Delta}{(\Delta^2+v^2p^2)^{3/2}}g^2sF(1,0) 
\end{equation}

To investigate transport coefficients, we use Eq(\ref{A17}) and Eq(\ref{A39})
\begin{equation}
\mathcal{G}_{ijk}=\sum_{\alpha}\int [dp](\mathcal{E}_{i}v_{j}-\mathcal{E}_{j}v_{i})p_{k}f_{0}
\end{equation}
 
\[\mathcal{G}_{xyy}=-\mathcal{G}_{yxy}=\sum_{\alpha}\int [dp]\mathcal{E}_{x}v_{y}p_{y}f_{0}\]
\[\mathcal{G}_{xyx}=-\mathcal{G}_{yxx}=0\]

\begin{equation}
    \mathcal{G}_{xyy}=\frac{v^2\Delta g^2sF}{m}\int \frac{d^2p}{(2\pi)^2}\frac{p_y^2}{(\Delta^2+v^2p^2)^{3/2}}f_0
\end{equation}
\[\mathcal{G}_{xyy}=\frac{v^2\Delta g^2sF}{m}\int \frac{dp}{(2\pi)^2}\frac{p^3}{(\Delta^2+v^2p^2)^{3/2}}\theta(\frac{p^2}{2m}+\Delta)\int_{0}^{2\pi} sin(\phi)^2d\phi\]

\begin{equation}
    \mathcal{G}_{xyy}=\frac{g^2s F}{4\pi }\big[-2\Delta+\frac{2\Delta^2+v^2P_{F}^2}{\sqrt{v^2P_{F}^2+\Delta^2}}\big]
\end{equation}

where
\begin{equation}
    P_{F}=\sqrt{2m|\Delta|}
\end{equation}
Also we can investigate non-linear transport in this approximation 
\begin{equation}
    G_{ijk}=\sum_{\alpha}\int [dp](\mathcal{E}_{i}v_{j}-\mathcal{E}_{j}v_{i})\frac{\partial f_0}{\partial p_k}\,,
\end{equation}
where the following terms are zero because of the Hamiltonian's symmetry ($k_x\rightarrow-k_x$) 
\[G_{xyy}=-G_{yxy}=\sum_{\alpha}\int [dp]\mathcal{E}_{x}v_{y}\frac{\partial f_0}{\partial p_y}\]
\[G_{xyx}=-G_{yxx}=0\]
and the non-zero coefficient is 
\begin{equation}
    G_{xyy}=2 g^2 s F\frac{P_{F}^2}{(\Delta^2+v^2P_{F}^2)^{3/2}}\,.
\end{equation}





\end{document}